\documentclass[american,aps,prl,superscriptaddress,reprint]{revtex4-1}
\usepackage[T1]{fontenc}
\usepackage[utf8]{inputenc}
\usepackage{color}
\usepackage{babel}
\usepackage{verbatim}
\usepackage{amstext}
\usepackage{graphicx}
\usepackage[unicode=true,pdfusetitle,
 bookmarks=true,bookmarksnumbered=false,bookmarksopen=false,
 breaklinks=true,pdfborder={0 0 0},pdfborderstyle={},backref=false,colorlinks=true]
 {hyperref}
\hypersetup{
 allcolors=blue}
\begin{document}
\title{\emph{M} center in 4\emph{H}-SiC is a carbon self-interstitial}
\author{J. Coutinho}
\email{jose.coutinho@ua.pt}

\address{I3N, Department of Physics, University of Aveiro, Campus Santiago,
3810-193 Aveiro, Portugal}
\author{J. D. Gouveia}
\altaffiliation{Current address: CICECO, Department of Chemistry, University of Aveiro, Campus Santiago, 3810-193 Aveiro, Portugal}

\address{I3N, Department of Physics, University of Aveiro, Campus Santiago,
3810-193 Aveiro, Portugal}
\author{T. Makino}
\address{Takasaki Advanced Radiation Research Institute, National Institutes
for Quantum and Radiological Science and Technology, 1233 Watanuki,
Takasaki, Gunma 370-1292, Japan}
\author{T. Ohshima}
\address{Takasaki Advanced Radiation Research Institute, National Institutes
for Quantum and Radiological Science and Technology, 1233 Watanuki,
Takasaki, Gunma 370-1292, Japan}
\author{Ž. Pastuović}
\address{Centre for Accelerator Science, Australian Nuclear Science and Technology
Organisation, 1 New Illawarra Rd, Lucas Heights, NSW 2234, Australia}
\author{L. Bakrač}
\address{Ruđer Bošković Institute, Bijenic\'{k}a 54, 10000 Zagreb, Croatia}
\author{T. Brodar}
\address{Ruđer Bošković Institute, Bijenic\'{k}a 54, 10000 Zagreb, Croatia}
\author{I. Capan}
\address{Ruđer Bošković Institute, Bijenic\'{k}a 54, 10000 Zagreb, Croatia}
\begin{abstract}
The list of semiconductor materials with spectroscopically fingerprinted
self-interstitials is very short. $M$ center in $4H$-SiC, a bistable
defect responsible for a family of electron traps, has been deprived
of a model which could unveil its real importance for almost two decades.
Using advanced first-principles calculations and junction spectroscopy,
we demonstrate that the properties of M, including bistability, annealing,
reconfiguration kinetics, and electronic levels, match those of the
carbon self-interstitial. {[}\emph{Pre-print published in Physical
Review B 103, L180102 (2021)}; DOI:\href{https://doi.org/10.1103/PhysRevB.103.L180102}{10.1103/PhysRevB.103.L180102}{]}
\end{abstract}

\maketitle
\noindent 

\noindent %

The identification of self-interstitials in technological crystalline
materials constitutes a rare event with profound repercussions on
several fields. Together with vacancies, they form a fundamental couple
playing a central role in many properties and processes, including
mass transport, crystal growth, doping and countless solid-state reactions.

\noindent %

Silicon carbide, in particular its $4H$ polytype ($4H$-SiC), is
a mainstream wide-gap semiconductor for power electronics \citep{Kimoto2014,She2017}
and a host for some of the most promising defects for quantum technologies
\citep{Weber2010,Koehl2011,Ivady2019,Castelletto2020}. In n-type
material, while both Si and C vacancies have well established spectroscopic
signals accompanied by detailed models \citep{Son2012,Trinh2013,Capan2018,Son2019,Defo2018,Udvarhelyi2020,Bathen2019b},
little is known about the interstitials, apart from indirect findings
from their interaction with vacancies \citep{Hiyoshi2009,Loevlie2011}
and some conjectural assignments. Deep Level Transient Spectroscopy
(DLTS) peaks EH$_{1/3}$ \citep{Hemmingsson1997,Alfieri2020}, M \citep{Martin2004,Nielsen2005}
and EB \citep{Beyer2011}, were to some extent related to carbon interstitials,
but when the data are confronted to existing models \citep{Bockstedte2003,Gali2003,Kobayashi2019},
we only obtain a partial match. This contrasts with p-type SiC, where
the connection of electron paramagnetic resonance signals T5 \citep{Itoh1997},
EI1 and EI3 \citep{Son2001} to carbon self-interstitials finds support
from first-principles calculations of hyperfine splitting constants,
zero-field coupling coefficients \citep{Petrenko2002} and $g$-tensor
elements \citep{Gerstmann2010}. %
{} Of course, having reliable fingerprints of self-interstitials in
$4H$-SiC, and particularly in n-type material which offers better
doping yield than p-type $4H$-SiC, would improve our ability to control
many defect engineering processes. For instance, an efficient production
of quantum technological defects in n-type $4H$-SiC (mostly vacancy
related and invariably introduced by ion-implantation or irradiation
techniques), strongly depends on our ability to control their annihilation
on capturing self-interstitials during subsequent thermal treatments.
This can only be achieved if we are able to follow the defects involved.

\noindent %

Below we demonstrate that the carbon self-interstitial (C$_{\textrm{i}}$)
is responsible for a family of DLTS traps, attributed nearly two decades
ago to a defect labeled ‘M’. They show up in irradiated n-type material
after moderate annealing ($T\sim200~^{\circ}\textrm{C}$) \citep{Martin2004,Nielsen2005},
including after irradiation with low-energy electrons (200~keV) \citep{Beyer2011},
implying a relation to a carbon sublattice defect \citep{Lefevre2009}.

The $M$ center is a bistable defect whose configuration is bias dependent.
Spectrum labeled ‘A’, evidencing traps M$_{1}$ ($E_{\textrm{c}}-0.42$~eV),
and M$_{3}$ ($\sim E_{\textrm{c}}-0.83$~eV) is obtained when the
sample is cooled from room temperature under reverse bias. Conversely,
when the DLTS scan is preceded by a gentle annealing at $T>140~^{\circ}\textrm{C}$
without bias, spectrum ‘B’ appears, consisting of a single emission
M$_{2}$ ($E_{\textrm{c}}-0.63$~eV) \citep{Martin2004,Nielsen2005}.
The spectra can be cycled without loss of signal amplitude by repeatedly
(un)biasing/annealing the sample. Since the M-peaks overlap with other
prominent traps in irradiated $4H$-SiC, namely Z$_{1/2}$ \citep{Hemmingsson1998}
and S$_{1/2}$ \citep{David2004}, the analysis must be carried out
from differential spectra.

By applying isothermal heat treatments, it was possible to extract
activation energies for the conversion between A and B configurations.
In unbiased samples the B$^{=}$ ground state builds up at the expense
of metastable A$^{=}$ with an activation energy of 1.4~eV. The superscripts
refer to the double minus charge state of the defect. Conversely,
in reverse biased samples, the A$^{-}$ state recovers from metastable
B$^{-}$ with an activation energy of 0.9~eV \citep{Martin2004}.

\noindent %

The reconfiguration within the neutral state could not be monitored
because the B signal consisted of M$_{2}$ only ($\textrm{B}^{=}\rightarrow\textrm{B}^{-}+e^{-}$).
It was argued that above room temperature, a fast conversion from
$\textrm{B}^{-}$ to $\textrm{A}^{-}$ frustrated the observation
of an expected M$_{4}$ peak related to $\textrm{B}^{-}\rightarrow\textrm{B}^{0}+e^{-}$,
effectively channeling any second emission through $\textrm{A}^{-}\rightarrow\textrm{A}^{0}+e^{-}$
\citep{Nielsen2005}.

\noindent %

Besides complying with the above observations, we show that C$_{\textrm{i}}$
features a rich set of properties that meet those of the $M$ center,
such as charge states and annealing. The model, developed using accurate
range-separated hybrid density functional calculations combined with
junction spectroscopy, includes a detailed configuration coordinate
diagram (CCD).

\noindent %

The electronic structure calculations were based on the planewave/projector-augmented
wave formalism \citep{Blochl1994} using the Vienna Ab initio Simulation
Package \citep{Kresse1996,Kresse1996b}, where planewaves with a cut-off
energy of 420~eV described the Kohn-Sham wave functions. Total energies
of stable and saddle-point structures were evaluated using the density
functional proposed by Heyd, Scuseria, and Ernzerhof (HSE06) \citep{Heyd2003}.
Defects were inserted in 400-atom $4H$-SiC hexagonal supercells whose
Brillouin zones (BZ) were integrated at $\mathbf{k}=\Gamma$. Forces
on atoms were calculated within a semilocal approximation to the exchange-correlation
interactions \citep{Perdew1996} and a BZ sampling grid of $2\times2\times2$
$\mathbf{k}$-points. This allowed for the investigation of stable
defect structures, transformation, and migration mechanisms. For the
mechanisms, we used the climbing image nudged elastic band (NEB) method
\citep{Henkelman2000}. All forces were kept below 0.01~eV/Å. Formation
energies as a function of the Fermi level were found by the usual
method \citep{Freysoldt2014}. Spurious periodic effects were mitigated
by adding a correction to the energies of charged supercells \citep{Freysoldt2009}.
For further details and testing, see Refs.~\citep{Bathen2019a,Coutinho2021}.

\noindent %

Experiments were carried out on Schottky diodes fabricated on 25-$\mu\textrm{m}$-thick
n-type $4H$-SiC layers (nitrogen doping up to $4.5\times10^{14}\,\textrm{cm}^{-3}$).
They were pattern-implanted through their nickel Schottky contacts
with 2-MeV He ions with a fluence of $10^{9}\,\textrm{cm}^{-2}$.
Conventional DLTS measurements were carried out in the temperature
range 100-450~K. Reverse voltage, pulse voltage and pulse width were
$V_{\textrm{R}}=-4$~V , $V_{\textrm{P}}=0$~V , and $t_{\textrm{P}}=10$~ms,
respectively. These conditions correspond to a depletion width of
1.7-3.5~$\mu\textrm{m}$, safely avoiding the ion projected range
(4.8~$\mu\textrm{m}$). In order to stabilize the spectra and monitor
structural transformations during the measurements, we applied isothermal
DLTS \citep{Tokuda1979,Tokuda2006}. Here, 20-min-long capacitance
transients were acquired at a sampling rate of 60~kHz, keeping the
temperature of the samples within $\pm0.05$~K from a set-point in
the range 280-340~K. Before quenching the diodes to the measurement
temperature, the $M$ centers were either set to configuration A or
B by applying a $-30$~V reverse bias annealing at 340~K for 20~min,
or annealing at 450~K for 20~min without bias, respectively.

\noindent %

\begin{figure*}
\includegraphics[width=16cm]{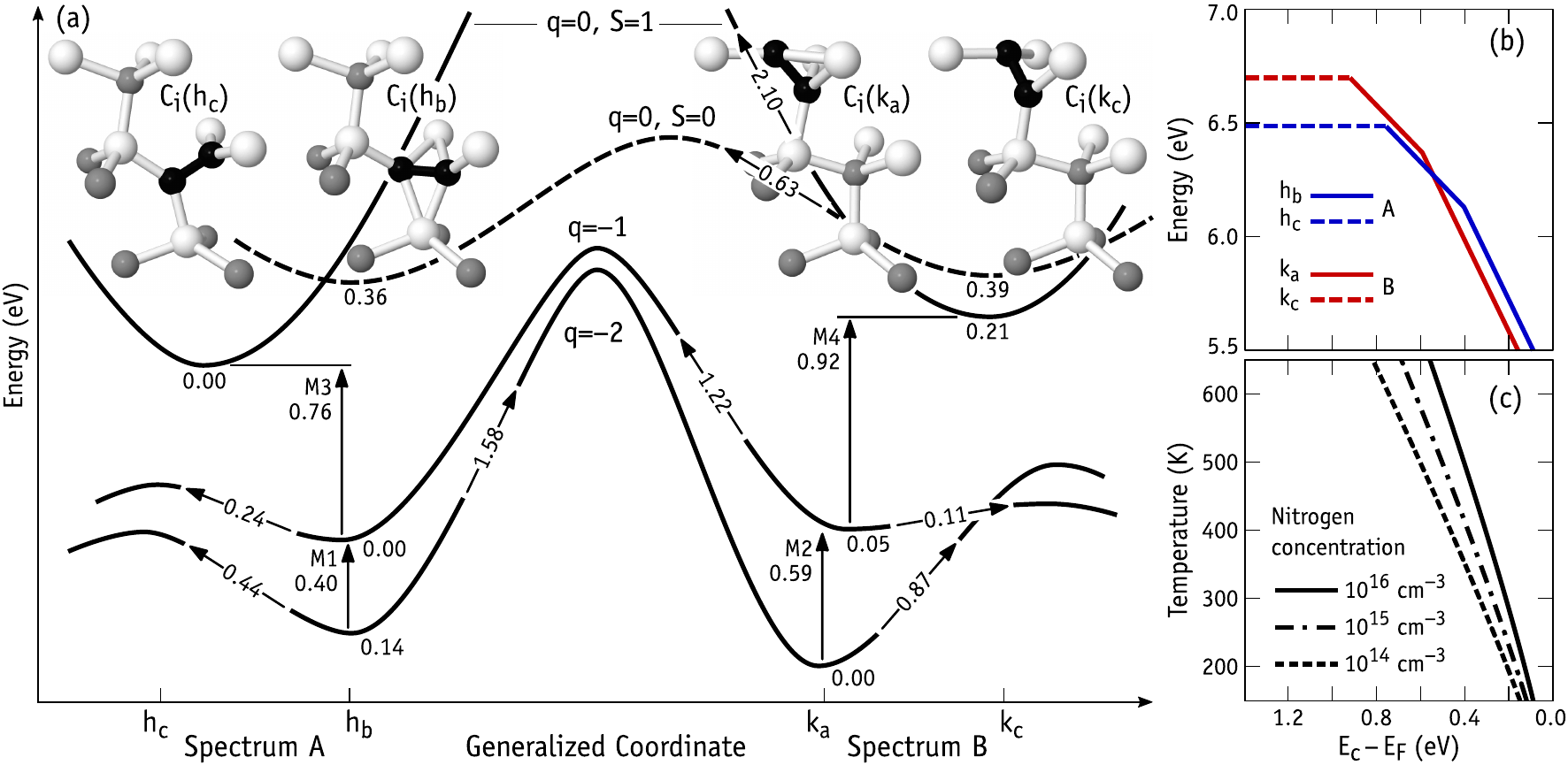}

\caption{\label{fig1}(a) Atomistic structures and CCD for the C$_{\textrm{i}}$
defect in n-type $4H$-SiC for charge states $q=0$, $-1$ and $-2$.
Si and C neighbors are shown in white and gray, respectively. The
C-dumbbell atoms are shown in black. Energies indicated below each
potential minimum are relative to the ground state of that charge
state. Energy barriers are relative to the structure of departure.
(b) Formation energy diagram of C$_{\textrm{i}}$ in n-type $4H$-SiC.
Energies of configurations A and B are shown in blue and red, respectively.
Dashed lines stand for spin-1 states. (c) Temperature dependence of
the Fermi level in n-type $4H$-SiC for three values of nitrogen concentration.}
\end{figure*}

The geometry and electronic state of C$_{\textrm{i}}$ in $4H$-SiC
strongly depends on its sublattice location and net charge, $q$.
We found four conspicuous configurations {[}see Fig.~\ref{fig1}(a){]},
some of which are already known \citep{Bockstedte2003,Gali2003,Kobayashi2019}.
We refer to them by a sublattice site ($k$ or $h$) and an index
letter (a, b or c) that stands for the alignment of the C-C dimer
{[}highlighted in black in Fig.~\ref{fig1}(a){]}.

Like the analogous defect in diamond \citep{Hunt2000}, neutral C$_{\textrm{i}}$
adopts a spin-1 state with three-fold coordinated C atoms. Two stable
structures, namely $\textrm{C}_{\textrm{i}}^{0}(h_{\textrm{c}})$
and $\textrm{C}_{\textrm{i}}^{0}(k_{\textrm{c}})$, are assigned to
the potential minima labeled with ‘$q=0$, $S=1$’, represented by
a solid line in the upper part of the CCD of Fig.~\ref{fig1}(a).
Diamagnetic S=0 states are metastable, and they were found with $h_{\textrm{b}}$
and $k_{\textrm{c}}$ structures (dashed line in the CCD). Energies
indicated below the potential minima are relative to the ground state
of the respective charge state. The defect can trap up to two electrons
via overcoordination of the dumbbell atoms. Negatively charged $h_{\textrm{b}}$
and $k_{\textrm{a}}$ (with $q=-1$ and $q=-2$) are now stable while
other geometries become unstable. NEB calculations show that $h_{\textrm{c}}$
and $k_{\textrm{c}}$ structures are saddle points for rotation of
$k_{\textrm{a}}$ and $h_{\textrm{b}}$ around the $\langle0001\rangle$
crystalline axis. The respective energy barriers are indicated in
the CCD over the arrows that seemingly follow the potential curves.
Note that $\textrm{C}_{\textrm{i}}^{-}(k)$ and $\textrm{C}_{\textrm{i}}^{-}(h)$
have rotation barriers as low as 0.11 and 0.24~eV. They are expected
to show an effective trigonal symmetry due to thermal motion, even
at cryogenic temperatures.

\noindent 

\noindent %

Figure~\ref{fig1}(b) shows the formation energy of C$_{\textrm{i}}$
in $4H$-SiC as a function of the Fermi level with respect to the
conduction band bottom, $E_{\textrm{c}}-E_{\textrm{F}}$. It was calculated
by assuming a C-rich crystal with the carbon chemical potential obtained
from bulk diamond. For the sake of clarity, line styles are connected
to specific geometries (see legend) and the range of $E_{\textrm{F}}$
values span the upper half of the gap only. Besides a charge state-dependent
structure, C$_{\textrm{i}}$ finds its most favorable sublattice site
depending on the location of the Fermi level. In n-type material (without
external bias), the most stable state is $\textrm{C}_{\textrm{i}}^{=}(k_{\textrm{a}})$,
whereas if the Fermi level is brought toward midgap (reverse bias
conditions), the ground state is $\textrm{C}_{\textrm{i}}^{0}(h_{\textrm{c}})$.
This behavior closely follows that of the $M$ center and its spectra
A and B, hereafter assigned to $\textrm{C}_{\textrm{i}}(h)$ and $\textrm{C}_{\textrm{i}}(k)$,
respectively.

\noindent %

Although the charge of defects is not directly accessible by capacitance
measurements, the lack of a Poole-Frenkel effect for the M$_{\textrm{1-3}}$
emissions indicates that they involve acceptor transitions \citep{Nielsen2005}.
Directly measured capture cross-sections of the shallower M$_{1}$
and M$_{2}$ traps were about $5\times10^{-17}\,\textrm{cm}^{2}$,
suggesting $(=/-)$ transitions. The cross-section of M$_{3}$ was
two orders of magnitude larger, consistent with M$_{1}$ and M$_{3}$
being two consecutive electron emissions from the same structure (spectrum
A). M$_{2}$ plus a conjectured (undetected) M$_{4}$ peak would account
for the analogous transitions in spectrum B. Hence, the type and number
of transitions conform with the assignment of C$_{\textrm{i}}$ to
M.

\noindent %

The $M$ center anneals out in the temperature range 580-640~K according
to first order kinetics at a rate of $1.5\times10^{13}\exp(-2\,\textrm{eV}/k_{\textrm{B}}T)\,\textrm{s}^{-1}$,
suggesting a dissociation mechanism or capture by an abundant and
close trap \citep{Nielsen2005}. Figure~\ref{fig1}(c) gives the
calculated Fermi energy in $4H$-SiC as a function of temperature
for three different nitrogen doping levels. The horizontal axis matches
that of the formation energy diagram in Fig.~\ref{fig1}(b). Considering
the doping concentration $1\textrm{-}2\times10^{15}\,\textrm{cm}^{-3}$
of the samples in Ref.~\citep{Nielsen2005}, we conclude that the
annealing temperature range coincides with the Fermi level window
for which C$_{\textrm{i}}$ is in the single negative charge state.
From NEB calculations we find that the barriers for migration of $\textrm{C}_{\textrm{i}}^{-}$
along crystalline basal and axial directions are 1.8 and 2.2 eV, respectively,
the latter being the highest of two jumps between consecutive sublattice
sites, and close to the measured activation energy for the annealing
of the $M$ center \citep{Nielsen2005}.

\noindent %

Let us now compare the charge state transition levels of C$_{\textrm{i}}$
with those of the $M$ center. The calculated figures are shown in
Fig.~\ref{fig1}(a). The energies next to the vertical arrows (in
eV) stand for trap depths with respect to the conduction band bottom.
Accordingly, $\textrm{C}{}_{\textrm{i}}(h)$ has second and first
acceptor transitions calculated at $E_{\textrm{c}}-0.40$~eV and
$E_{\textrm{c}}-0.76$~eV, very close to the measured transitions
of spectrum A, namely M$_{1}$ ($E_{\textrm{c}}-0.42$~eV) and M$_{3}$
($E_{\textrm{c}}-0.83$~eV). Analogously, for $\textrm{C}{}_{\textrm{i}}(k)$
we find second and first acceptor transitions at $E_{\textrm{c}}-0.59$~eV
and $E_{\textrm{c}}-0.92$~eV. The former agrees well with M$_{2}$
($E_{\textrm{c}}-0.63$~eV) from spectrum B, while the latter anticipates
the location of M$_{4}$. The calculated levels neglect any temperature
effects. They differ from the measured activation energies by a capture
barrier and an entropy contribution, usually $\sim\!0.1$~eV. Hence,
the agreement obtained is considered very good.

Besides the above, the calculated $\textrm{C}_{\textrm{i}}(h)\leftrightarrow\textrm{C}_{\textrm{i}}(k)$
conversion barriers further confirm that the $M$ center in $4H$-SiC
is a carbon interstitial. Like the carbon vacancy \citep{Bathen2019a},
in order to travel between two equivalent lattice sites along the
main crystalline axis, C$_{\textrm{i}}$ must perform two types of
jumps between consecutive sublattice configurations. The jump with
lower barrier is represented in Fig.~\ref{fig1}(a) and separates
$\textrm{C}{}_{\textrm{i}}(h)$ from $\textrm{C}{}_{\textrm{i}}(k)$
structures. Calculated barrier heights of 1.58 and 1.22~eV were obtained
for forward and backward jumps $\textrm{C}_{\textrm{i}}^{=}(h_{\textrm{b}})\rightarrow\textrm{C}_{\textrm{i}}^{=}(k_{\textrm{a}})$
and $\textrm{C}_{\textrm{i}}^{-}(h_{\textrm{b}})\leftarrow\textrm{C}_{\textrm{i}}^{-}(k_{\textrm{a}})$,
respectively. Both mechanisms are exothermal, and the barriers are
rather close to the $\textrm{A}\rightarrow\textrm{B}$ and $\textrm{A}\leftarrow\textrm{B}$
conversion activation energies of 1.4 and 0.9~eV, measured without
and with bias, respectively \citep{Martin2004}.

\noindent %

\begin{figure*}
\includegraphics[width=16cm]{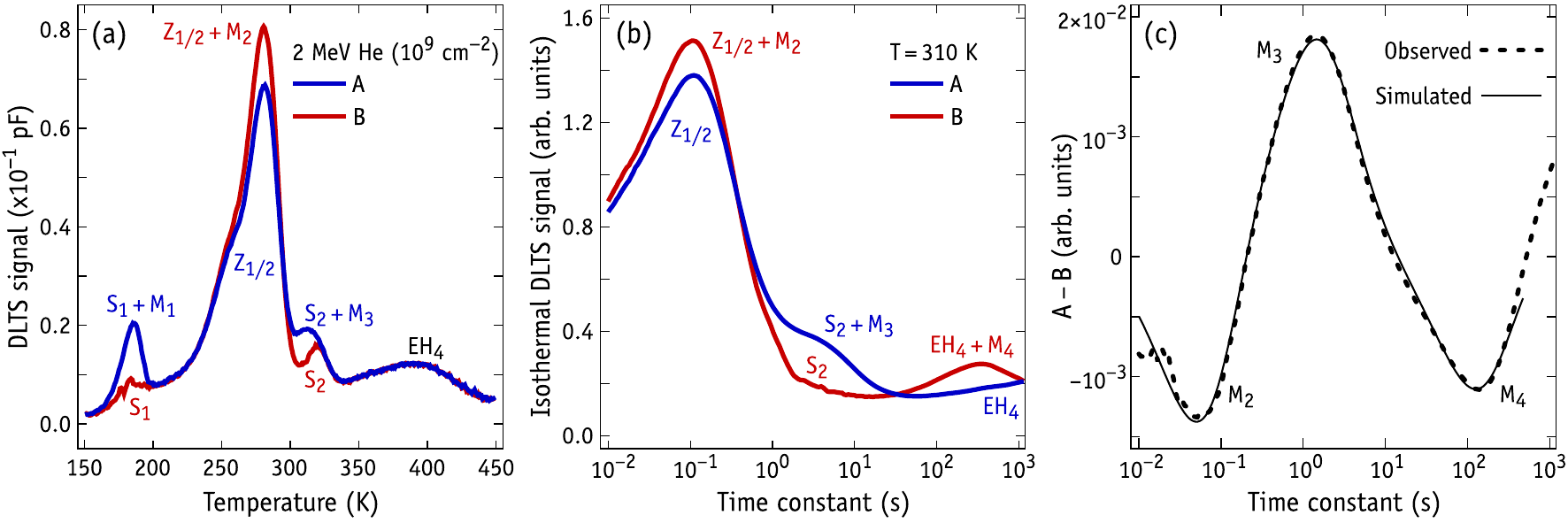}

\caption{\label{fig2}(a) DLTS spectra of 2 MeV He implanted $4H$-SiC SBD
showing two configurations of the $M$ center (A and B) along with
other (labeled) deep traps. (b) Isothermal DLTS signals of the $M$
center in configurations A (blue) and B (red) at 310~K. M$_{2}$,
M$_{3}$, and M$_{4}$ overlap with Z$_{1/2}$, S$_{2}$ and EH$_{4}$,
respectively. (c) Differential signal (dotted curve) of the isothermal
spectra shown in (b). The thin solid line represents the simulated
signal from M$_{2}$, M$_{3}$, and M$_{4}$ deep traps obtained by
nonlinear fitting. See text for experimental details.}
\end{figure*}

A decisive piece of evidence in the identification of the $M$ center
would be a direct observation of the M$_{4}$ transition accompanied
by M$_{2}$. Figure~\ref{fig2}(a) shows DLTS spectra of a $4H$-SiC
n-type sample implanted with 2~MeV He ions, measured after annealing
at 340~K under reverse bias (spectrum A) and 450~K without bias
(spectrum B). The latter is dominated by the Z$_{1/2}$ peak (carbon
vacancy) \citep{Hemmingsson1998,Son2012}, with relatively smaller
contributions from S$_{1}$/S$_{2}$ (silicon vacancy) \citep{David2004,Bathen2019b},
and a less understood broad feature known as EH$_{4}$, recently connected
to a superposition of different alignments of anisotropic silicon
vacancy structures commonly referred to as carbon-antisite-vacancy
pairs \citep{Karsthof2020}.

\noindent 

It is only when we compare spectrum B with A that the presence of
the bistable $M$ center becomes evident. The prominent peak at about
280~K in spectrum B is actually a superposition of Z$_{1/2}$ with
M$_{2}$, the latter being converted to M$_{1}$ and M$_{3}$ in spectrum
A (overlapping S$_{1}$ and S$_{2}$, respectively). As in Ref.~\citep{Nielsen2005}
we find that above room temperature spectra A and B become identical
as a result of the $\textrm{B}\rightarrow\textrm{A}$ transformation.
This means that we do not have access to $M$ center related transitions
deeper than M$_{3}$ (as it is expected for the location of M$_{4}$)
by means of conventional DLTS.

Figure~\ref{fig2}(b) depicts two isothermal DLTS spectra acquired
sequentially at $T=310$~K. Spectrum B was acquired first, after
pre-annealing the sample at 450~K with no bias applied. During the
20-min recording time at $T=310$~K, a large fraction of $M$ centers
was converted from configuration B to A. This could be confirmed from
the loss of intensity of M$_{2}$ during subsequent identical measurements.
Spectrum A was then recorded after warming up the sample at $T=340$~K
under reverse bias ($-30$~V ) in order to reset all $M$ centers
to the A form. The isothermal signals have contributions from three
groups of transients: A fast group with time constant $\tau\sim10^{-1}$~s
(Z$_{1/2}$ and M$_{2}$), an intermediate group with decay time of
a few seconds (S$_{2}$ and M$_{3}$), and a slow group in the range
$\tau\sim10^{2}\textrm{-}10^{3}$~s (EH$_{4}$ and a new peak labeled
M$_{4}$). The M$_{4}$ peak is clearly shown in the differential
spectrum of Fig.~\ref{fig2}(c). It has the same behavior of M$_{2}$
on annealing/biasing, and corresponds to a trap concentration identical
to that of M$_{1}$, M$_{2}$, and M$_{3}$ ($7\textrm{-}8\times10^{11}\,\textrm{cm}^{-3}$).

From a fit of a $T^{2}$-corrected Arrhenius function to the M$_{4}$
data, we obtain an activation energy for electron emission of $E_{\textrm{a}}=0.86\pm0.02$~eV
and an extrapolated capture cross-section $\sigma_{\textrm{a}}=4\times10^{-15}\,\textrm{cm}^{2}$
which is coherent with a first acceptor transition. Although we could
not directly measure the capture cross-section of M$_{4}$, the agreement
with the calculated $\textrm{C}_{\textrm{i}}^{-}(k_{\textrm{a}})\rightarrow\textrm{C}_{\textrm{i}}^{0}(k_{\textrm{c}})+e^{-}$
transition energy of 0.92~eV is striking. Also notable is the agreement
shown in Fig.~\ref{fig2}(c) between the isothermal data (dotted
line) and its theoretical counterpart (solid line) \citep{Tokuda1979},
obtained via non-linear fitting constrained by the measured emission
rates of M$_{2}$, M$_{3}$ and M$_{4}$.

\noindent %

Unfortunately, the measurements cannot discern whether M$_{4}$ is
an evidence of (i) $\textrm{B}^{-}\rightarrow\textrm{B}^{0}+e^{-}$
followed by $\textrm{B}^{0}\rightarrow\textrm{A}^{0}$ or (ii) $\textrm{B}^{-}\rightarrow\textrm{A}^{-}$
immediately followed by $\textrm{A}^{-}\rightarrow\textrm{A}^{0}+e^{-}$.
The second and first steps of (i) and (ii), respectively, are necessary
to explain the gradual conversion of B into A during the isothermal
transients. The calculations account for $\textrm{B}^{0}\rightarrow\textrm{A}^{0}$
in (i) if we consider the spin-flipping transition shown as a dashed
line in the CCD. On the other hand, the first step of route (ii) has
a measured activation energy of 0.9~eV, very close to the M$_{4}$
emission energy, and this could reflect the rate-limiting process
toward $\textrm{A}^{0}$. Regarding the C$_{\textrm{i}}$ model, although
being consistent with both alternatives, this issue certainly calls
for further clarification.

Another aspect that deserves a closer look, concerns the annealing
mechanism of the M center. The observed first order kinetics suggests
an annealing rate that depends on its own concentration only. A possible
explanation stems from the indication that C$_{\textrm{i}}$ defects
are most likely negatively charged during annealing, thus being Coulomb-attracted
by abundant nitrogen donors. The products of such reaction could be
the EB centers reported by Beyer \emph{et~al.} \citep{Beyer2011}.

The list of alternative defects, other than C$_{\textrm{i}}$, that
could be related to the $M$ center is not long. A Si displacement
is unlikely, given that M was observed in material irradiated with
electrons whose energy was below the necessary displacement threshold
\citep{Beyer2011}. Further, we know that the Si self-interstitial
is not an acceptor and has levels considerably deeper than those of
M \citep{Coutinho2021}.

A stronger contender is a carbon Frenkel pair. Carbon vacancies and
self-interstitials have electron traps deeper than $\sim E_{\textrm{c}}-0.4$~eV.
In n-type material they are both negatively charged, and Coulomb interactions
between pairs with varying distances should lead to a strong dispersion
of their respective trap depths. This effect is observed in DLTS spectra
of as irradiated samples, but does not conform with the relatively
clean M peaks, which are only visible after a gentle ($T\sim200\,^{\circ}\textrm{C}$)
annealing \citep{Doyle1998,David2004}. We investigated the stability
of close Frenkel pairs and found a family of markedly low-energy structures.
Their formation energy is in the range 5.5-6.5~eV but the annihilation
barrier is about 0.1-0.2~eV and therefore could not survive even
at room temperature.

\noindent %

Based on first-principles calculations and junction spectroscopy,
we provide substantial and credible arguments that allow us to conclude
that the $M$ center in $4H$-SiC is a carbon self-interstitial. The
observation of a new peak, labeled M$_{4}$ and possibly connected
to a previously anticipated acceptor transition, provides an important
piece of evidence in the connection between M and the interstitial.
The model developed is summarized in Fig~\ref{fig1}(a) in the form
of a CCD. It incorporates the observed features of M, including charge
states, bistability, annealing, reconfiguration kinetics and electronic
transition levels. The $M$ center peaks are either ``hidden'' by
the prominent Z$_{1/2}$ signal (carbon vacancy) or by S$_{1/2}$
(silicon vacancy) and EH$_{1/3}$ (unidentified). While this may explain
the difficulty in the identification of C$_{\textrm{i}}$ in n-type
$4H$-SiC, it certainly has undermined the interpretation of defect
evolution by means of DLTS, in particular during annealing. The identification
of the $M$ center in $4H$-SiC as the C$_{\textrm{i}}$ defect will
contribute to a better understanding of defect-related processes,
ranging from self-diffusion to the activation and migration of dopants
and impurities introduced by ion-implantation.

\section*{Acknowledgments}

This work was supported by the NATO SPS Programme through Project
G5674. J.C. thanks the FCT in Portugal for support through Projects
UIDB/50025/2020, UIDP/50025/2020, and CPCA/A0/7277/2020 (Advanced
Computing Project using the Oblivion supercomputer). J.D.G. acknowledges
the support of I3N through Grant BPD-11(5017/2018). T.O. thanks Dr.
Hidekazu Tsuchida and Dr. Norihiro Hoshino at CRIEPI for growing the
$4H$-SiC epitaxial layers. Z.P. acknowledges the financial support
of the Australian Government to the CAS of ANSTO through the NCRIS.

\bibliographystyle{apsrev4-1}

%

\end{document}